\documentclass{article}

\usepackage{PRIMEarxiv}
\usepackage{authblk}
\usepackage[utf8]{inputenc} 
\usepackage[T1]{fontenc}    
\usepackage{hyperref}       
\usepackage{url}            
\usepackage{booktabs}       
\usepackage{amsfonts}       
\usepackage{nicefrac}       
\usepackage{microtype}      
\usepackage{lipsum}
\usepackage{fancyhdr}       
\usepackage{graphicx}       
\graphicspath{{media/}}     

\usepackage{cite}

\usepackage{hyperref}
\hypersetup{
  colorlinks=true,
  linkcolor=black,
  citecolor=black,
  urlcolor=black
}
\title{Denoising Simulated Low-Field MRI (70mT) using Denoising Autoencoders (DAE) and Cycle-Consistent Generative Adversarial Networks (Cycle-GAN)

\thanks{2023 International Society of Magnetic Resonance in Medicine.  Toronto, Canada,  June 2-9.  Abstract Number 1764}
}

\author[1-4]{Fernando Vega }
\author[1-4]{Abdoljalil Addeh}
\author[1-4]{M. Ethan MacDonald}

\affil[1]{Department of Biomedical Engineering, Schulich School of Engineering, University of Calgary, Calgary, AB, Canada}
\affil[2]{Department of Electrical \& Software Engineering, Schulich School of Engineering, \protect\\University of Calgary, Calgary, AB, Canada}
\affil[3]{Hotchkiss Brain Institute, Cumming School of Medicine, University of Calgary, Calgary, AB, Canada}
\affil[4]{Department of Radiology, Cumming School of Medicine, University of Calgary, Calgary, AB, Canada}

\begin{document}
\maketitle
\section*{Synopsis}
In this work, a denoising Cycle-GAN (Cycle Consistent Generative Adversarial Network) is implemented to yield high-field, high resolution, high signal-to-noise ratio (SNR) Magnetic Resonance Imaging (MRI) images from simulated low-field, low resolution, low SNR MRI images.  Resampling and additive Rician noise were used to simulate low-field MRI. Images were utilized to train a Denoising Autoencoder (DAE) and a Cycle-GAN, with paired and unpaired cases.  Both networks were evaluated using SSIM and PSNR image quality metrics. This work demonstrates the use of a generative deep learning model that can outperform classical DAEs to improve low-field MRI images and does not require image pairs. 
\section*{Introduction}
Over the last few decades there has been an increasing use of magnetic resonance imaging (MRI) as it provides hundreds of contrast modes and is minimally invasive \cite{RN9, RN8}.  It is known that higher spatial resolution and SNR-efficiency can be achieved with higher field strength \cite{RN10}. However, as the field strength increases, so does the cost \cite{doi:10.1148/radiol.2020192084}. Low-Field MRI scanners are less expensive (\textasciitilde20x less expensive than 3T over 10 years), have much lower energy consumption (\textasciitilde60x less electricity) \cite{RN11}, reduce the energy absorption in the subject, and do not require expensive liquid helium \cite{RN6}; however, the trade-off is lower resolution and lower SNR-efficiency \cite{RN3}. 

\noindent Previous work aimed to improve the resolution and SNR-efficiency by implementing machine learning techniques such as a Denoising Autoencoder (DAE) \cite{RN12, RN21} that uses Convolutional Neural Networks (CNN) \cite{RN13}. However, using this architecture requires the images to be paired and aligned.  Performing registration in noisy images is prone to error necessitating a technique that does not need images to be paired or registered. This led us to use a Cycle Consistent Generative Adversarial Network (Cycle-GAN) \cite{RN1} as an improvement over classical DAEs. 

\noindent Cycle-GAN architecture is also based on CNNs, it uses four networks: two generators and two discriminators, where one generator produces synthetic denoised images that are fed to a second generator that generates the original noisy image. One discriminator is assigned to each generator to predict if the generated images are real or synthetic \cite{RN14}. Using this approach, GAN architectures excel at generating synthetic images with a high degree of similarity to the real ones \cite{RN15}. 

\noindent In this work, a 3D Cycle-GAN was implemented using unpaired 3T MRI images and low-field simulated MRI images.  The model was evaluated with unseen images and reported the Structural Similarity Index (SSIM) \cite{RN17} and Peak Signal-to-Noise Ratio (PSNR) \cite{RN18} as performance metrics.  These results are compared with the performance of DAEs.

\section*{Method}
100 T1-weighted MRI images were used from Open Access Series of Imaging Studies (OASIS-3) \cite{RN7} database (3T MRI images with a resolution of 1mm $\times$1mm$\times$1mm).  Then low-field MRI images were synthesised to have a resolution of 1.5mm$\times$1.5mm$\times$1.5mm and added Rician noise to emulate a low SNR of 70mT \cite{RN4}. 

\begin{figure}[p]
\includegraphics[width=\textwidth]{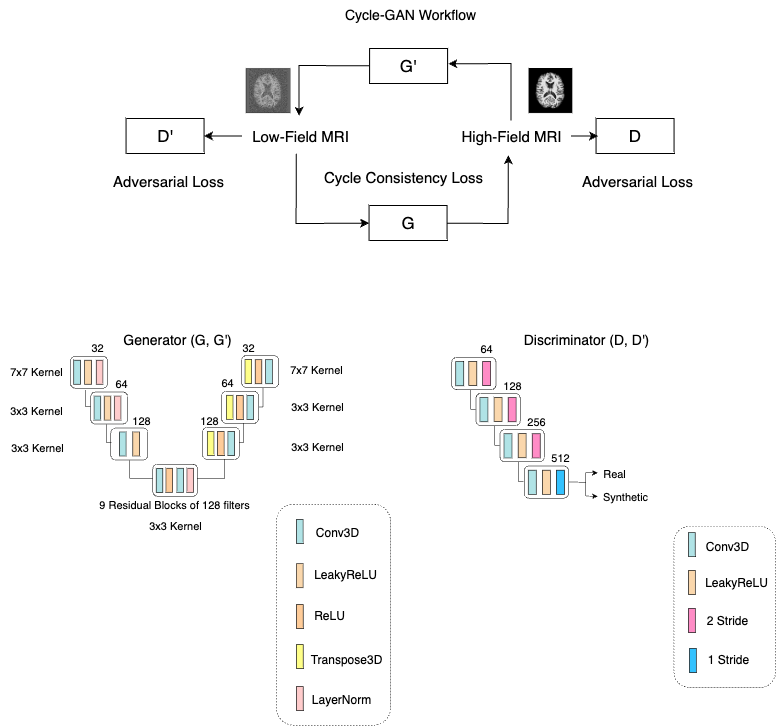}
\caption{Cycle-GAN workflow(top), it relies on 2 generators: G generates a high-field MRI from the low-field MRI, G’ generates the low-field MRI back from the generated high-field MRI to enable the model to deal with unpaired data. The output from both generators is sent to the respective discriminators which classify the generated images as real or synthetic. The generator (bottom-left) follows an encoder-decoder architecture, and the discriminator(bottom-right) is a classifier.} 
\label{cycle-arch}
\end{figure}

\noindent A 3D Cycle-GAN model was implemented using the MONAI deep learning framework [18], the model was fed with 100 high-field MRI images and 100 simulated low-field MRI images for 500 epochs following the architecture shown in \ref{cycle-arch}. This architecture has a total of 13 layers with 9 residual blocks that act as a bottleneck without any skip connection, as shown in Figure \ref{cycle-arch}. This architecture diverges from the standard U-net style followed in DAEs. 

\noindent Once the model was trained, it was evaluated with 100 unseen images and the results were compared with a DAE and evaluated using the SSIM and PSNR metrics.

\section*{Results}
\begin{figure}[h]
\includegraphics[width=\textwidth]{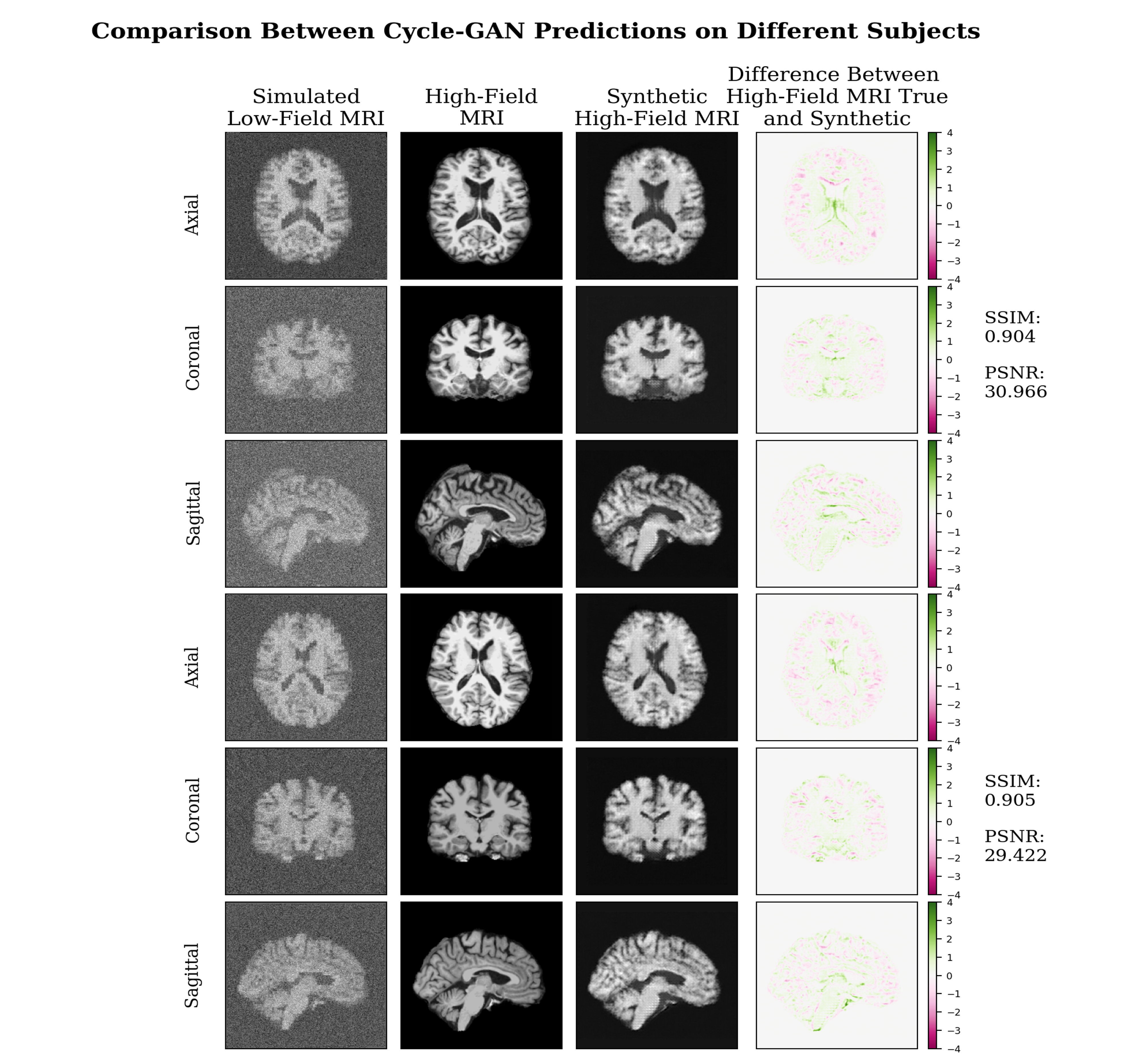}
\caption{Cycle-GAN High-Field MRI predictions panel: Columns left: Low-Field MRI, center-left: High-field MRI, center-right: Synthetic High-Field MRI generated by the model, right: Difference map between high-field MRI true and synthetic images. 2 subjects with planes from the 3D images. Based on the reported SSIM and PSNR model can produce synthetic images that have high degree of similarity with the true in both shape and contrast.} 
\label{Cycle-results}
\end{figure}

\begin{figure}[h]
\includegraphics[width=\textwidth]{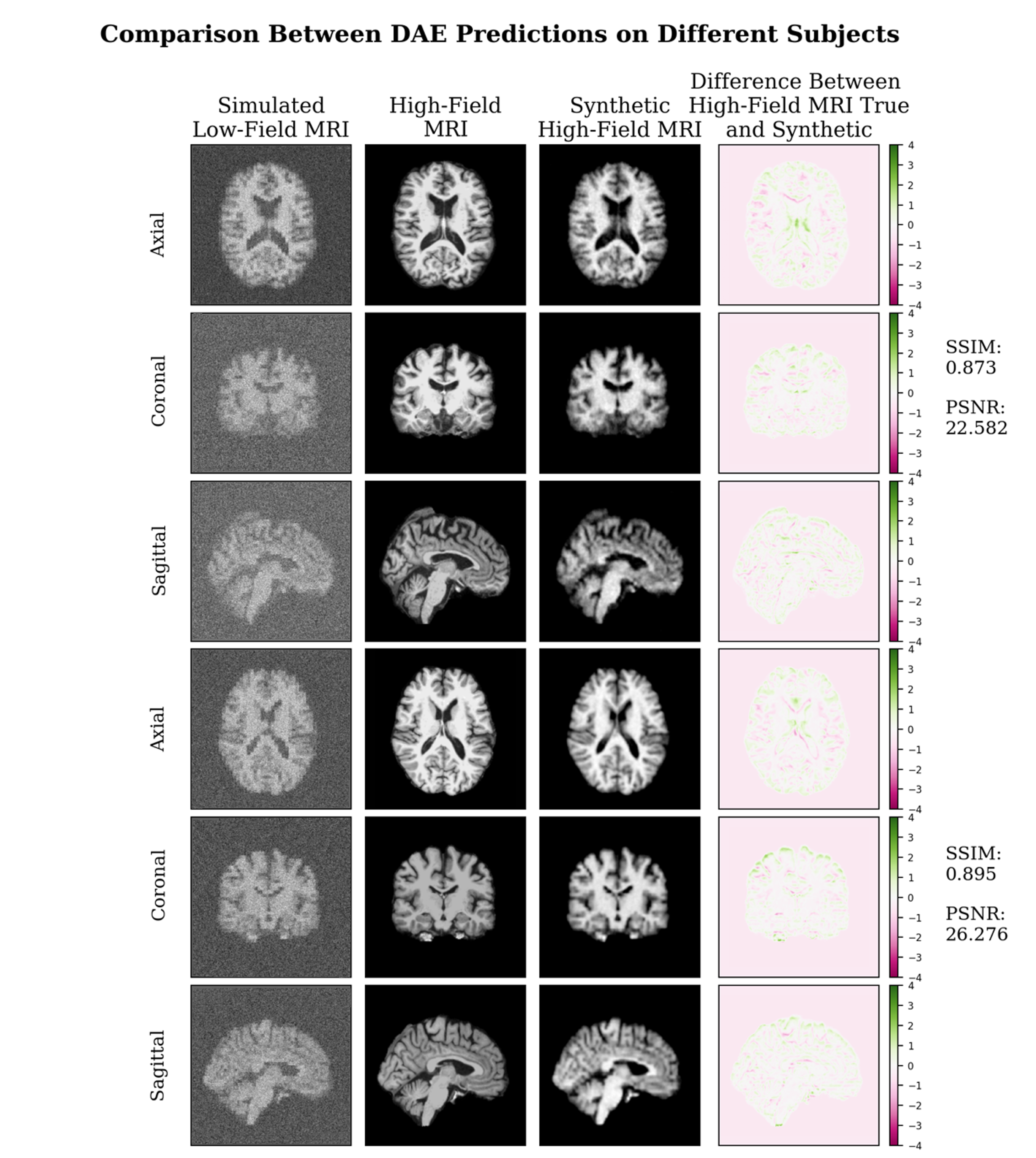}
\caption{DAE High-Field MRI predictions panel: Columns left: Low-Field MRI, center-left: High-field MRI, center-right: Synthetic High-Field MRI generated by the model, right: Difference map between high-field MRI true and synthetic images. 2 subjects with planes from the 3D images, DAE produces good High-Field MRI images, however, Cycle-GAN achieves higher SSIM and PSNR in the shown subjects. DAE performs good in the exclusive case where paired imaging is available.} 
\label{Dae-results}
\end{figure}

\begin{figure}[h]
\includegraphics[width=\textwidth]{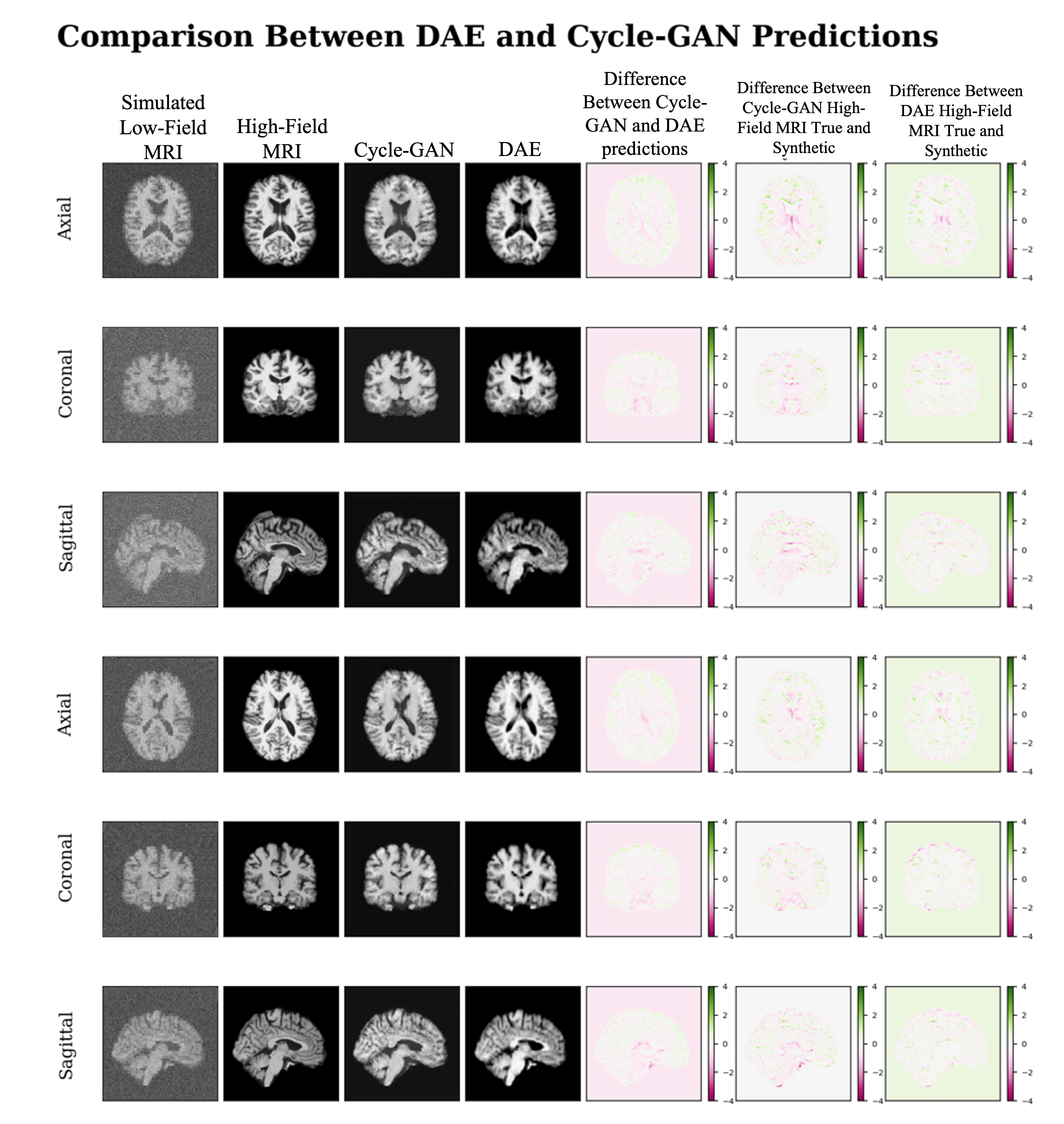}
\caption{Comparison between model DAE and Cycle-GAN panel, from left to right  first column: simulated Low-Field MRI, second column: High-Field MRI, third column:, Cycle-GAN synthetic High-Field MRI, fourth column: DAE synthetic High-Field MRI, fifth column: Difference map between Cycle-GAN and DAE synthetic High-Field MRI, sixth column: Difference map between Cycle-GAN predictions and High-Field MRI, seventh column: Difference map between DAE predictions and High-Field MRI.} 
\label{Dae-cycle}
\end{figure}

The results obtained can be seen in Figure \ref{Cycle-results}, where the synthetic images have a high degree of visual similarity with the true images based on the reported SSIM and PSNR, Figure \ref{Dae-results} shows the same subjects using a DAE. In Figure \ref{Dae-cycle}, the Cycle-GAN denoising model is compared with a DAE showing that the Cycle-GAN produces overall better images in terms of contrast and shape. 

\begin{figure}[h]
\includegraphics[width=\textwidth]{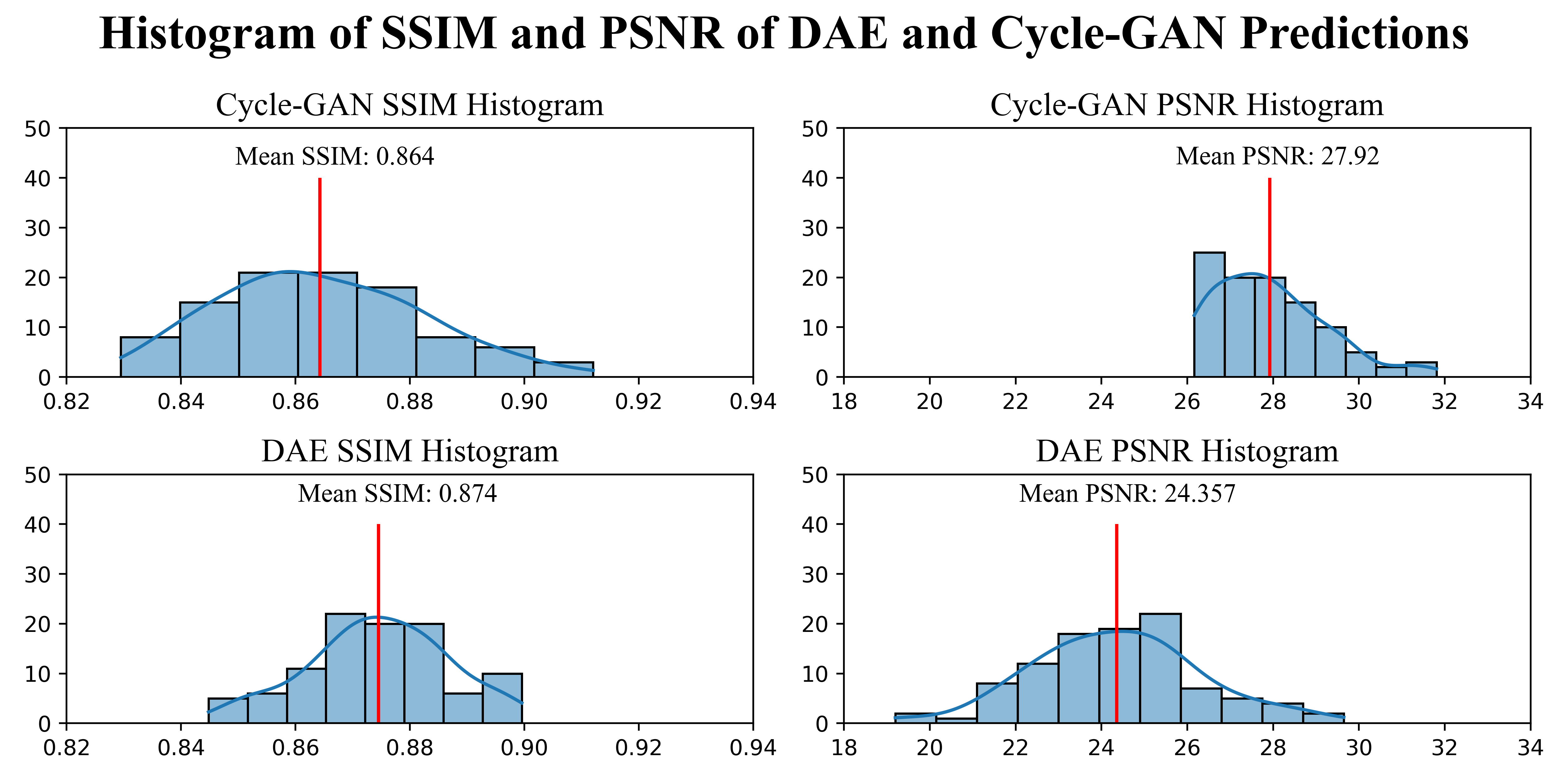}
\caption{Histogram of SSIM and PSNR in both Cycle-GAN (top) and DAE (bottom). DAE has a higher mean SSIM, however, Cycle-GAN has a higher PSNR. Since PSNR is more sensitive to detect noise, this indicates that Cycle-GAN tends to produce cleaner images. It is expected a high SSIM in DAE as paired images were used to compare both models and DAE excels with paired images. } 
\label{Histograms}
\end{figure}

\noindent The metrics tested in the cohort of unseen images show that the Cycle-GAN model is able to produce high quality synthetic denoised images as shown in Figure \ref{Histograms} with a mean PSNR 14.62\% higher than the DAE. The DAE scored 1.15\% higher in SSIM compared to the Cycle-GAN.However, the PSNR is a more sensible measure to compare noise between images than the SSIM.

\section*{Discussion}
This work demonstrates a pipeline that can produce similar or better estimations than classical DAE in low-field simulated images. The results are encouraging as it proves that low-field MRI images can be used to generate images with the same quality as a high-field MRI without the need of paired data.  In future work, we propose to address the limitations of this project. One, being the use of simulated low-field data that needs to be replaced with empirically gathered low-field data to produce a representative model.   Another limitation in this simulation is that we do not consider T1, T2 differences at different field strengths. 

\noindent This work is a major advance as it shows that the Cycle-GAN performs better than the DAE and does not require image pairs in training.

\section*{Acknowledgements:}
The authors would like to thank the University of Calgary, in particular the Schulich School of Engineering and Departments of Biomedical Engineering and Electrical and Software Engineering; the Cumming School of Medicine and the Departments of Radiology and Clinical Neurosciences; as well as the Hotchkiss Brain Institute, Research Computing Services and the Digital Alliance of Canada for providing resources. The authors would like to thank the Open Access of Imaging Studies Team for making the data available. FV – is funded in part through the Alberta Graduate Excellence Scholarship.  JA – is funded in part from a graduate scholarship from the Natural Sciences and Engineering Research Council Brain Create. MEM acknowledges support from Start-up funding at UCalgary and a Natural Sciences and Engineering Research Council Discovery Grant (RGPIN-03552) and Early Career Researcher Supplement (DGECR-00124).\\

\bibliographystyle{unsrt}
\bibliography{Denoiser_GAN_ref.bib}

\end{document}